\documentclass[amsmath,amssymb,amsbsy,reprint,prl,preprintnumbers,showpacs,superscriptaddress]{revtex4-2}
\usepackage{graphicx,color}
\usepackage{dcolumn}
\usepackage{bm}
\usepackage{braket}
\usepackage{mathtools}
\usepackage{ulem}
\usepackage[breaklinks,colorlinks=true,linkcolor=blue,urlcolor=blue,citecolor=blue]{hyperref}
\usepackage{times}
\usepackage{physics}
\usepackage{latexsym}
\usepackage{amsmath, amssymb}
\usepackage{mathtools}
\usepackage{multirow}

\newcommand{\be}{\begin{eqnarray}}
\newcommand{\ee}{\end{eqnarray}}

\begin{document}
\title{Edge bits in average symmetry protected topological mixed state}

\date{\today}
\author{Yoshihito Kuno} 
\affiliation{Graduate School of Engineering science, Akita University, Akita 010-8502, Japan}


\begin{abstract} 
Edge bit in an average symmetry protected topological (ASPT) mixed state is studied. The state is protected by one strong $Z_2$ and one weak (average) $Z_2$ symmetries. As analogous objects of pure symmetry protected topological (SPT) states, the ASPT possesses edge bits. In particular, the analogous operator response exists, that is, symmetry fractionalization. The fractionalization preserves the presence of the ASPT in the bulk, and the fractionalized edge operators acting on the edge bits of the ASPT. 
In this work, based on the cluster model and by employing Choi mapping, we discuss generic features of the edge bits and numerically clarify the behavior of the edge bits and their robustness for varying decoherence and perturbative interactions. By using an operator-space mutual information (OSMI), we track the flow of quantum correlations between the two edges. Remarkably, even in the ASPT regime, a finite portion of the initial edge-to-edge correlation survives. 
\end{abstract}


\maketitle
{\it Introduction.---}
Symmetry-protected topological (SPT) state is a hallmark in the discovery of non-trivial quantum many-body phases \cite{Pollman2010,XieChen2011,Pollman2012,Senthil2015}. 
If we impose a symmetry on a system, non-trivial many-body phases can be defined. The state cannot be characterized by the Landau paradigm.  
The variation of SPT is rich. The group cohomology approach classifies SPTs, giving the richness of the SPT \cite{XieChen2011,Chen2013,Chen2014}.
As a specific property, the SPT setting on a system under boundaries exhibits edge modes rooted in the presence of the bulk SPT; it is called ``bulk-edge correspondence" \cite {Hatsugai1993,XieChen2011,Schuch2011}.
Especially, the protecting symmetry to the SPT generates edge-mode operators, exhibiting anomalous anticommutation relations, t'Hooft anomaly \cite{Else2014}. This generation of the edge-mode operator is called ``symmetry fractionalization'' \cite{XieChen2011,Pollman2012,Verresen2017}.

Recently, the notion of the SPT phase has been extended into open quantum systems \cite{Nieuwenburg2014,groot2022}. As a specific extension, an average symmetry protected topological phase (ASPT) has been proposed \cite{ma2023}. The ASPT is protected by the extended notion of symmetry defined on mixed states, namely weak symmetry \cite{Buca_2012,groot2022}.
The ASPT is intrinsically different from pure-state conventional SPTs. 
The ASPT features a symmetry-protection mechanism fundamentally different from that of pure-state SPTs. Notably, it is robust against symmetry-preserving decoherence and other open-system perturbations \cite{Shah2025}.
Currently, some mixed state ASPT under decoherences were clarified \cite{Lee2025,Guo-and-Ashida2024,min2024,Ando2024,Guo2024_1,Guo2024_2}.  
These studies elucidated the varieties of ASPTs and of realizing intrinsically new SPT phases with no counterpart in the pure-state framework. Especially, the group cohomology classification has been extended to classify the ASPTs \cite{Ma_PRXQuantum6}. 
So far, most of the analytical and numerical studies of the ASPT focus on the bulk and its bulk ASPT order. For example, a bulk extended string order to characterize the bulk SPT \cite{Lee2025,Guo2024_1,Guo2024_2,Shah2025}. At the level of the abstract notion, the ASPT also has bulk-edge correspondence \cite{Ma_PRXQuantum6}. The ASPT is expected to exhibit an edge bit. However, the concrete description and demonstrations are still lacking.
This work studies the edge bits of the ASPT by considering an extended cluster model \cite{Raussendorf2001,Son2011,Son2012} under decoherences. 
Based on the Choi isomorphism approach \cite{Choi1975,lee2023,Lee2025}, the edge bit in the cluster model under decoherences is discussed by especially focusing on the symmetry-fractionalization. We discuss a clear operator description elucidating the behavior of the edge bit in the cluster ASPT, the bulk physics of which has been studied \cite{Ma_PRXQuantum6,Lee2025,Shah2025}. Moreover, we numerically observe the behavior of the symmetry fractionalization of the ASPT and also clarify the correlation property between edge bits. We explicitly show that the quantum correlation of the cluster SPT with the Bell-pair edges survives even when the system turns into the ASPT. 

{\it System.---} This work focuses on an extended cluster Hamiltonian \cite{Raussendorf2001}
\begin{eqnarray}
H_{c} 
=-\sum^{L-3}_{j=0}Z_{j}X_{j+1}Z_{j+2}+J_{xx}\sum^{L-2}_{j=0}X_{j}X_{j+1},
\end{eqnarray}
where $L$ is the system size, taking an odd number, and an open boundary condition is imposed.
The system possesses $Z_2 \times Z_2$ symmetry\cite{Son2012}, the generators of which are global spin-flip operators $W\equiv \prod_{j\in even}X_{j}$ and $S\equiv \prod_{j\in odd}X_{j}$, respectively. 
For $0\leq J_{xx}<1$, the ground state is the cluster SPT state protected by the $Z_2\times Z_2$ symmetry. For $J_{xx}> 1$, the spontaneous symmetry breaking (SSB) phase appears.

Under the open boundary and $J_{xx}=0$, the explicit logical operators are known, $Z_{0(L)}$ and $X_{0(L-1)}Z_{1(L-2)}$ with $\{Z_{0(L)},X_{0(L-1)}Z_{1(L-2)}\}=0$ \cite{Verresen2017}. If we take $z$-polarized spin basis in the edge space, the four degenerate ground states are described as $|\Psi_{\alpha\beta}\rangle$ with $(Z_{0(L-1)})|\Psi_{\alpha\beta}\rangle=(-1)^{\alpha(\beta)}|\Psi_{\alpha\beta}\rangle$.
In this work, then we apply an even-site decoherence to the state of the model denoted by $\rho$, which is given by 
\begin{eqnarray}
\mathcal{E}^{Z}_{e}[\rho]=\prod_{j\in even}\mathcal{E}^{Z}_{j}[\rho],\;\:
\mathcal{E}^{Z}_{j}[\rho] = \biggr[(1-p_{z})\rho+p_{z}Z_{j}\rho Z_{j}\biggl]. 
\end{eqnarray}
$p_z$ is the parameter controlling the strength of the decoherence taking $0\leq p_{z}\leq 1/2$.

For the analysis of the density matrix $\rho\in \mathcal{H}$, we use the doubled Hilbert space formalism, namely, Choi mapping \cite{Choi1975,JAMIOLKOWSKI1972}, where the target Hilbert space is doubled as $\mathcal{H}_{u}\otimes \mathcal{H}_{\ell}$, where the subscripts $u$ and $\ell$ denote the upper and lower Hilbert spaces corresponding to ket and bra states of mixed state density matrix, respectively.
The density matrix $\rho$ is vectorized called Choi map, $\rho \longrightarrow |\rho\rangle\rangle$ as $|\rho\rangle\rangle\equiv \frac{1}{\sqrt{\dim[\rho]}}\sum_{k}|k\rangle_{u}\otimes \rho|k\rangle_{\ell}$, where $\{|k\rangle_{u} \}$ and $\{|k\rangle_{\ell} \}$ is an orthonormal set of bases in the Hilbert spaces $\mathcal{H}_{u}$ and $\mathcal{H}_{\ell}$. 
Then, the Choi map of a pure state $|\psi_0\rangle$ is described as $\rho_0 \equiv|\psi_0\rangle \langle \psi_0|\longrightarrow |\rho_0\rangle\rangle\equiv |\psi^*_0\rangle|\psi_0\rangle$.
In this formalism the decoherence $\mathcal{E}^{Z}_{e}$ can be treated as a non-unitary operator to the Choi map, $\hat{\mathcal{E}}^{Z}_{e}=\prod_{j\in even}[(1-p_{z})\hat{I}_{j,u}^* \otimes \hat{I}_{j,\ell} +p_{z}Z_{j,u}^*\otimes Z_{j,\ell}]$, where the labels $(j,u)$ and $(j,\ell)$ on the operators mean acting on the site $j$ residing on the upper(bra) space $\mathcal{H}_{u}$ and lower(ket) space $\mathcal{H}_{\ell}$, these notation ``$u$'' and ``$\ell$'' for operators are used in the following.
Then, a mixed Choi map state by $\mathcal{E}^{Z}_{e}$ is produced as 
$
|\rho^D\rangle\rangle\equiv\hat{\mathcal{E}}^{Z}_{e}|\rho_0\rangle\rangle$ \cite{lee2023,Grover2024,Lee2025}. 
The above decohered Choi map can be efficiently manipulated by matrix product state and the filtering operations, where the numerical way has been developed in the previous studies\cite{Haegeman2015,Orito2025,KOI2025_v2}. The definition of these symmetries is shown in Appendix A. 

{\it Matrix product operator picture.---} For $J_{xx}=0$ and $p_z=1/2$ limit, it is known that the cluster ASPT appears from the cluster SPT under the decoherence $\mathcal{E}^{Z}_{e}$  \cite{Ma_PRXQuantum6,Lee2025,Ma2024_double}. 
As for Ref.\cite{Shah2025}, 
the cluster ASPT can be described as a matrix product operator (MPO) \cite{Verstraete2004}
\begin{eqnarray}
&&\rho^c_{\alpha\beta}\equiv \sum_{{\bf z},{\bf{x}},{\bf z}',{\bf{x}'}}{}_{e}\langle \alpha|[A^{(z_0,z'_0)}B^{(x_1,x'_1)}\cdots\nonumber\\
&& B^{(x_{L-2},x'_{L-2})}A^{(z_{L-1},z'_{L-1})}]|\beta\rangle_e |({\bf z},{\bf{x}})\rangle \langle ({\bf z}',{\bf{x}}')|.
\label{fixed_MPO}
\end{eqnarray} 
Here $\alpha,\beta=0,1$ and $|0\rangle_e=[1,0]^t$ and $|1\rangle_e=[0,1]^t$, $A^{(z,z')}=[(\frac{1+z}{2})P^{z(v)}_{+}+(\frac{1-z}{2})P^{z(v)}_{-}]\delta_{z,z'}$, where $P^{z(v)}_{\pm }\equiv|0(1)\rangle_e {}_{e}\langle 0(1)|$ is a projector acting on the virtual spin residing on the virtual space. Similarly, $B^{(x,x')}=({X^{v}})^{(1-x)/2}\delta_{x,x'}$, where $X^{v}$ is a Pauli $X$ operator acting on the virtual spin. The physical basis of odd sites and even ones are different: On the odd, $x$-direction is used  and on the even, $z$-direction is used and then each entire basis is written by $|({\bf z},{\bf x})\rangle$ with ${\bf x}=(x_0,x_2,\cdots,x_L)$ and ${\bf z}=(z_1,z_3,\cdots,z_{L-1})$ (taking $x_i,z_j=\pm 1$).
Throughout this work, the Choi map is mainly employed. The above density matrix is represented by a pure state vector called the Choi map \cite{lee2023,Lee2025}. Then, the Choi map of the above MPO for the fixed point ASPT can be regarded as a pure MPS \cite{Ma_PRXQuantum6} without normalization as $\rho^{c}_{\alpha\beta}\to |\rho^c_{\alpha\beta}\rangle\rangle$.
The fixed point ASPT is protected by the weak symmetry $W$ and the strong symmetry $S$, where the notion of both weak and strong symmetries is introduced in \cite{Buca_2012,groot2022}. 
The bulk ASPT order has been characterized by the extended string order \cite{Lee2025,Guo2024_1,Guo2024_2,Shah2025}.

{\it Symmetry fractionalization and edge operators.---} In the MPO representation, $\rho^c_{\alpha\beta}$ [Eq.~(\ref{fixed_MPO})], the responses to Pauli operators acting on physical legs are already known as the pulling-through relations \cite{Shah2025}. 
These are described as 
\begin{eqnarray}
&&\includegraphics[width=0.95\linewidth]{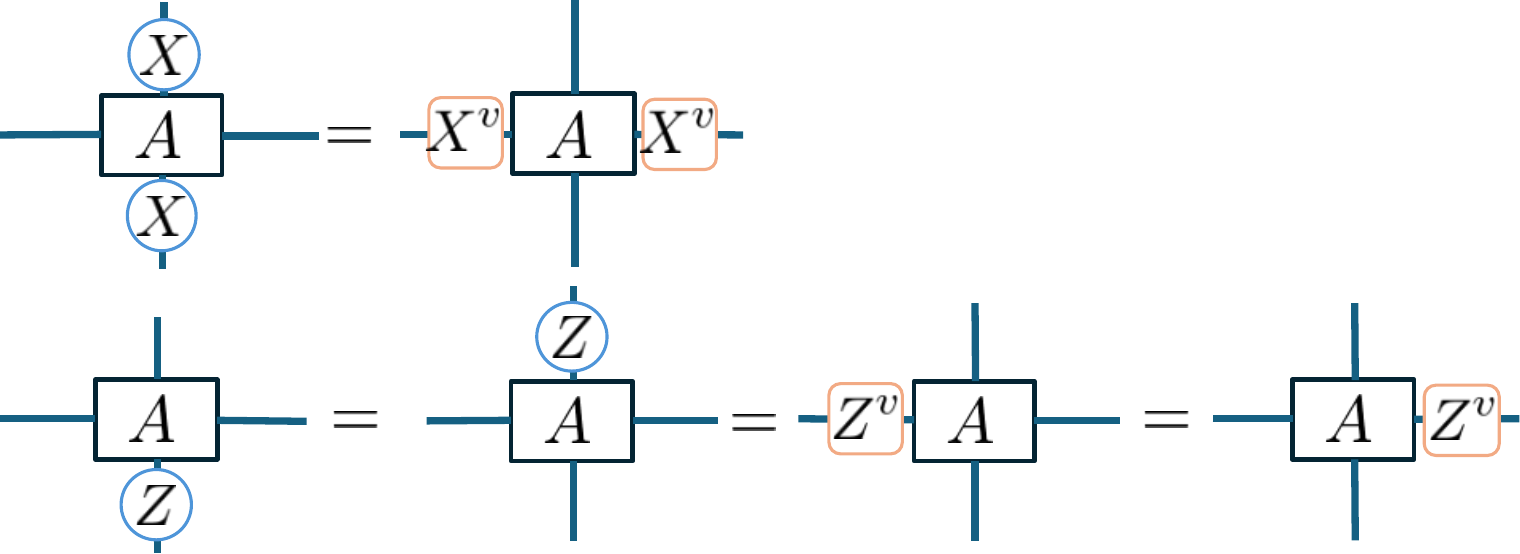}\nonumber\\
&&\includegraphics[width=0.75\linewidth]{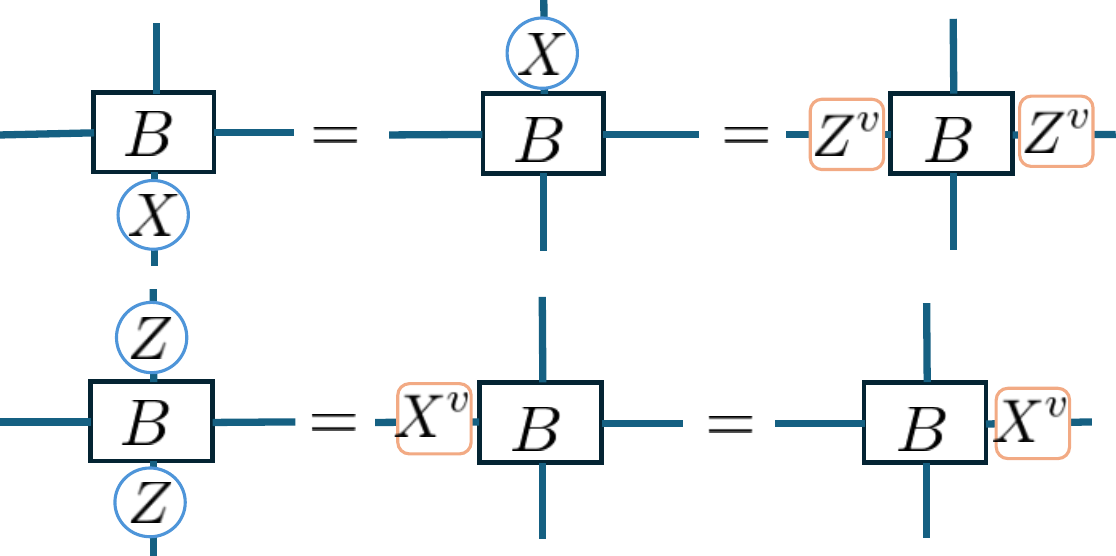}\label{eq:pullB}\nonumber
\end{eqnarray}
where the vertical lines are physical legs and the horizontal lines represent the virtual space. The application of local $Z$ or $X$ Pauli operators to the physical legs induces Pauli matrices $X^{v}$ and $Z^{v}$ on the virtual spins residing on the two-dimensional virtual space \cite{Gross2007}. 
This pulling-through relations elucidate the response of the application of strong $S$ and weak $W$ symmetries, giving non-trivial edge behavior of the ASPT. This is a symmetry fractionalization deduced as \cite{Shah2025,Ma_PRXQuantum6}
\begin{eqnarray}
&&(W^*_u\otimes W_{\ell})|\rho^c_{\alpha\beta}\rangle\rangle=(X^{v}_{L})(X^{v}_{R})|\rho^c_{\alpha\beta}\rangle\rangle\nonumber\\
&&=[(X_{0,u}Z_{1,u})(Z_{L-2,u}X_{L-1,u})]\times\nonumber\\
&&[(X_{0,\ell}Z_{1,\ell})(Z_{L-2,\ell}X_{L-1,\ell})]|\rho^c_{\alpha\beta}\rangle\rangle,\label{WSF_eq}\\
&&(S^*_u\otimes {\bf 1}_{\ell})|\rho^c_{\alpha\beta}\rangle\rangle=(Z^{v}_{L})(Z^{v}_{R})|\rho^c_{\alpha\beta}\rangle\rangle\nonumber\\
&&=(Z_{0,u}Z_{L-2,u})|\rho^c_{\alpha\beta}\rangle\rangle=(Z_{0,\ell}Z_{L-2,\ell})|\rho^c_{\alpha\beta}\rangle\rangle,
\end{eqnarray}
where $X^{v}_{L(R)}$ and $Z^{v}_{L(R)}$ are Pauli operators applying to the virtual space at edges, that is, acting on the virtual boundary state $|\alpha\rangle_{e}$, representing the edge states. 
From these fractionalization, physical edge operators for the fixed cluster ASPT can be regarded as $Z$ and $XZ$.
Although this mixed state formalism does not provide the quantum edge qubit, the physical edge operators $Z$ and $XZ$ act like a logical operator. The state $|\rho^c_{\alpha\beta}\rangle\rangle$ edge bit polarized as $Z_{0(L)}=(-1)^{\alpha(\beta)}$ since $\Tr[\rho^c_{\alpha\beta}Z_{0(L)}]=(-1)^{\alpha(\beta)}$. 
The edge operator $XZ$ in Eq.~(\ref{WSF_eq}) acts as the bit-flip operator since $\{Z_{0(L-1)},X_{0(L-1)}Z_{1(L-2)}\}=0$. Indeed, for a mixed state with edge bits,  $(X_0Z_1)\rho^{c}_{\alpha \beta}(X_0Z_1)^{\dagger}=\rho^c_{{\bar \alpha}\beta}$ with $\bar{\alpha}=(\alpha+1) \bmod 2$ [$(X_{0,u}Z_{1,u})(X_{0,\ell}Z_{1,\ell})|\rho^{c}_{\alpha \beta}\rangle\rangle=|\rho^c_{{\bar \alpha}\beta}\rangle\rangle$].
In addition, the edge bit polarized as $Z_{0(L-1)}=\pm 1$ at the fixed point does not affected by the decoherence $\mathcal{E}^{Z}_{e}$ since $\mathcal{E}^{Z}_{e}[Z_{0(L-1)}]=Z_{0(L-1)}$. 

In the framework of the Choi map, the ASPT state with edge bits $|\rho_{\alpha\beta}\rangle\rangle$ can be superposed as
$|\rho\rangle\rangle=\sum_{\alpha\beta}p_{\alpha\beta}|\rho_{\alpha\beta}\rangle\rangle$ with $p_{\alpha\beta}\geq 0$. 
This superposition of the Choi map indeed corresponds to a classical probabilistic mixture of the density matrix with different polarized edge bits,  $\rho=\sum_{\alpha,\beta=0,1}p_{\alpha\beta}\rho_{\alpha\beta}$.
One can consider the edge bits with a correlation between edges such as $|\rho\rangle\rangle=(1/2)[|\rho^c_{00}\rangle\rangle+|\rho^c_{11}\rangle\rangle]$. 
\begin{figure}[t]
\begin{center} 
\vspace{0.5cm}
\includegraphics[width=7.8cm]{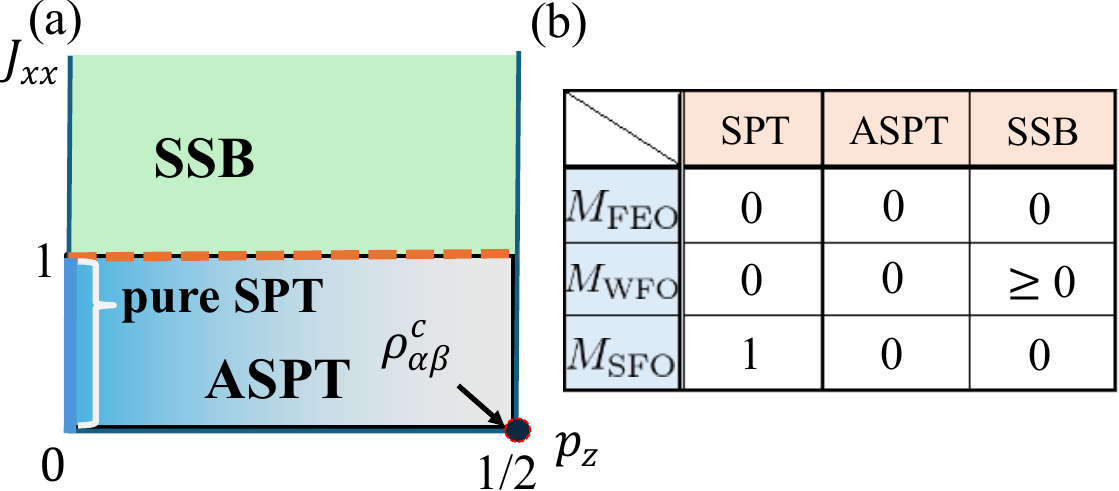}  
\end{center} 
\caption{(a) Schematic phase structure. There are three states in the system under the decoherence $\mathcal{E}^{Z}_{e}$ and the nearest-neighbor $XX$ interactions. (b) Table of behavior of the order parameters of the symmetry fractionalization. The patterns of values of the order parameters can characterize each phase from the viewpoint of the presence of edge operators.}
\label{Fig_PD}
\end{figure}

Up to this point, we discussed the properties of the edge bits of the ASPT under the fixed point ($p_z=1/2$, $J_{xx}=0$). We next consider how the edge bit behaves apart from the fixed point. As far as the weak $W$ and strong $S$ symmetries are preserved and any spontaneous symmetry breaking in the bulk does not occur, edge bits in the ASPT exist, although both forms of explicit edge bit operators $Z$ and $XZ$ and the MPO of the ASPT of Eq.~(\ref{fixed_MPO}) are deformed.  We expect that the symmetry fractionalizations are still preserved, which is generally described by
\begin{eqnarray}
(W^*_u\otimes W_{\ell})|\rho_{\alpha\beta}\rangle\rangle
&=&(Q^{(p)}_{L,u}Q^{(p)}_{R,u})(Q^{(p)}_{L,\ell}Q^{(p)}_{R,\ell})|\rho_{\alpha\beta}\rangle\rangle,\label{WSF_eq}\\
(S^*_u\otimes {\bf 1}_{\ell})|\rho_{\alpha\beta}\rangle\rangle
&=&(E^{(p)}_{L,u}E^{(p)}_{R,u})|\rho_{\alpha\beta}\rangle\rangle\nonumber\\
&=&(E^{(p)}_{L,\ell}E^{(p)}_{R,\ell})|\rho_{\alpha\beta}\rangle\rangle.
\end{eqnarray}
where $|\rho_{\alpha\beta}\rangle\rangle$ is a deformed Choi map from the fixed point ASPT, the operators $Q^{(p)}_{L(R),u(\ell)}$ and $E^{(p)}_{L(R),u(\ell)}$ are dressed logical operators deformed from  $X_{0(L-1),u(\ell)}Z_{1(L-2),u(\ell)}$ and $Z_{0(L-1),u(\ell)}$. These dressed operators are localized around edges. Also, the anticommutation relation $\{Q^{(p)}_{L(R),u(\ell)},E^{(p)}_{L(R),u(\ell)}\}=0$ is expected.

Herein, we show the way to verify the presence of the symmetry fractionalization, and from this fractionalization, we characterize the ASPT phase and its robustness. 
To this end, we introduce three order parameters to characterize the symmetry fractionalization expected above. The first one is a 
fixed-point edge-operator order parameter (FEO)
\begin{eqnarray}
M_{\rm FEO}\equiv\frac{\langle\langle \rho^{D}_{\alpha\beta}|(X_0Z_1)_{u}\otimes (X_0Z_1)_{\ell} |\rho^{D}_{\alpha\beta}\rangle\rangle}{\langle\langle \rho^{D}_{\alpha\beta}|\rho^{D}_{\alpha\beta}\rangle\rangle},
\end{eqnarray} 
corresponding to $\Tr[(X_0Z_1)\rho^D_{\alpha\beta}(X_0Z_1)\rho^D_{\alpha\beta}]/\Tr[(\rho^{D}_{\alpha\beta})^2]$. Here, $|\rho^{D}_{\alpha\beta}\rangle\rangle$ is a state generated by the initial pure SPT with edge bits and by the decoherence $\mathcal{E}^Z_e$, where $\alpha$ and $\beta$ are polarized label for the left and right edge bits such as $Z_{0(L-1)}=(-1)^{\alpha(\beta)}$ or $E^{(p)}_{L(R),u}=E^{(p)}_{L(R),\ell}=(-1)^{\alpha(\beta)}$ \cite{commute_bit_polarization}.
Without the SSB, this quantity takes zero if the state $|\rho^{D}_{\alpha\beta}\rangle\rangle$ correctly response to the fixed-point edge operator $XZ$ even though the genuine bit-flip operator is slightly different from the fixed ones. We expect that for the ASPT (and the pure SPT), $M_{\rm FEO}$ vanishes. For the SSB phase, since the $XX$ interaction anticommute to $XZ$ operator, $M_{\rm FEO}$ also vanishes. 
\begin{figure}[t]
\begin{center} 
\includegraphics[width=8.8cm]{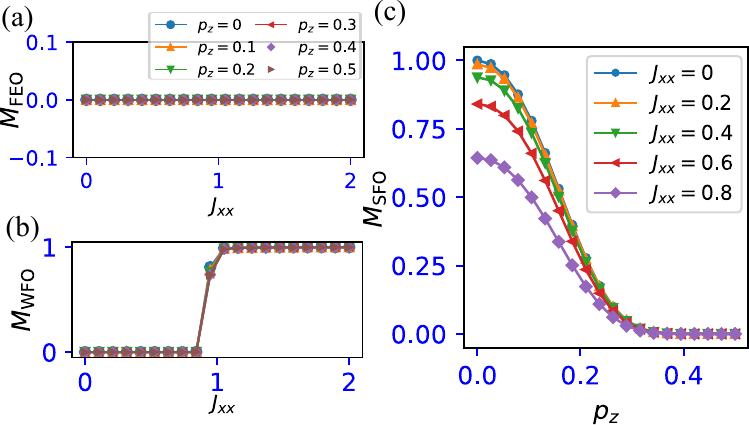}  
\end{center} 
\caption{Behaviors of the order parameters of the symmetry fractionalization. (a) $J_{xx}$-dependence of $M_{\rm FEO}$. (b) $J_{xx}$-dependence of $M_{\rm WFO}$. 
In (a) and (b), the data for different $p_z$'s are displayed. (c) $J_{xx}$-dependence of $M_{\rm SFO}$ for different $J_{xx}$'s.
$L=23$ ($23\times 2$-sites ladder). The system size dependence for (c) is shown in Appendix F.
}
\label{Fig_M}
\end{figure}
As the second quantity, we introduce a weak-symmetry fractionalization order parameter (WFO)
\begin{eqnarray}
M_{\rm WFO}\equiv\frac{\langle\langle \rho^D_{\alpha\beta}|W^*_{u}\otimes W_{\ell} |\rho^{D}_{\alpha\beta}\rangle\rangle}{\langle\langle \rho^D_{\alpha\beta}|\rho^D_{\alpha\beta}\rangle\rangle},
\end{eqnarray} 
corresponding to $\Tr[W\rho^D_{\alpha\beta} W \rho^D_{\alpha\beta}]/\Tr[(\rho^D_{\alpha\beta})^2]$.
This characterizes the presence of the symmetry fractionalization for the weak symmetry operation like Eq.~(\ref{WSF_eq}). If the state $|\rho^D_{\alpha\beta}\rangle\rangle$ possess the symmetry fractionalization, $W^*_{u}\otimes W_{\ell} |\rho^{D}_{\alpha\beta}\rangle\rangle=(Q^{(p)}_{L,u}Q^{(p)}_{R,u})(Q^{(p)}_{L,\ell}Q^{(p)}_{R,\ell})|\rho^{D}_{\alpha\beta}\rangle\rangle=|\rho^{D}_{{\bar \alpha}{\bar \beta}}\rangle\rangle$,  giving $M_{\rm WFO}=0$. Also, the breakdown of the fractionalization makes $M_{\rm WFO}$ finite. In particular, the SSB induced by the $XX$ interactions gives a ferromagnetic order, giving $M_{\rm WFO}=1$, and if we set the state of the ASPT on a periodic boundary, $M_{\rm WFO}=1$. As the third, a partially strong-symmetry fractionalization order parameter (SFO) for the symmetry $W$ is introduced as
\begin{eqnarray}
M_{\rm SFO}\equiv\frac{\langle\langle \rho^{D}_{\alpha\beta}|[(X_0Z_1)_{u}\otimes (Z_{L-2}X_{L-1})_{u}][W^*_{u}\otimes I_{\ell}]|\rho^{D}_{\alpha\beta}\rangle\rangle}{\langle\langle \rho^{D}_{\alpha\beta}|\rho^{D}_{\alpha\beta}\rangle\rangle}.\nonumber
\end{eqnarray} 
This value can characterize the presence of the symmetry fractionalization of the pure SPT. If the state $|\rho^{D}_{\alpha\beta}\rangle\rangle$ is in the pure SPT, then $W_{u}\otimes I_{\ell}|\rho^{D}_{\alpha\beta}\rangle\rangle\sim(X_0Z_1)_{u}\otimes (Z_{L-2}X_{L-1})_{u}|\rho^{D}_{\alpha\beta}\rangle\rangle$. Thus, for the pure SPT, even for this strong symmetry operation, the symmetry fractionalization of the pure cluster SPT occurs, then $M_{\rm SFO}$ takes $\sim 1$. On the other hand, if the ASPT appears, such a fractionalization does not occur, and for the SSB, it is also true, that $M_{\rm SFO}$ takes $\sim 0$. We summarize the behaviors for these three quantities in Fig.~\ref{Fig_PD}(b).

{\it Numerical observation for symmetry fractionalizations.---} We numerically observe these order parameters above for a decohered state $|\rho^D_{e}\rangle\rangle\equiv\hat{\mathcal{E}}^{Z}_{e}|\rho^0_{e}\rangle$, where $|\rho^0_{e}\rangle\rangle\equiv |\Psi^*_{e}\rangle|\Psi_{e}\rangle$ and $|\Psi_{e}\rangle$ is a pure cluster SPT with edge states. In our numerics, we employ the TeNPy library \cite{10.21468/SciPostPhysLectNotes.5,10.21468/SciPostPhysCodeb.41,10.21468/SciPostPhysCodeb.41-r1.0}. We first prepare the initial state $|\rho^0_{e}\rangle\rangle$ by using the density matrix renormalization group algorithm (DMRG) on the ladder geometry (Each upper and lower chain corresponds to the upper and lower Hilbert spaces). We set maximum bond dimension $D=100$-$200$ and truncate the singular value less than $\mathcal{O}(10^{-6})$, and the energy convergence of the iterative DMRG sweeping is $\Delta E < \mathcal{O}(10^{-4})$ to obtain the initial MPS ground state $|\rho^0_{e}\rangle\rangle$. 
Then, the filtering operation of $\hat{\mathcal{E}}^{Z}_{e}$ to the state $|\rho^0_{e}\rangle\rangle$ is efficiently carried out numerically as the practical way is shown in \cite{Orito2025,KOI2025_v2}.

In the preparation of the initial pure state, we prepare the double cluster SPT states with edge bits much close to $Z_{L-1(L-1)}=+1$ (exact for $J_{xx}=0$), denoted by $|\rho^0_{e}\rangle\rangle=|\rho^0_{00}\rangle\rangle$. Practically this is carried out by adding a very small boundary magnetic $Z$ field on both edge sites on both Hilbert space. Then, we calculate the decohered state $\hat{\mathcal{E}}^Z_e|\rho^0_{e,00}\rangle\equiv |\rho^{D}_{00}\rangle\rangle$ for various cases and observe the above order parameters.

Figure \ref{Fig_M} shows the numerical results for the behaviors of the three quantities, $M_{\rm FEO}$, $M_{\rm WFO}$, and $M_{\rm SFO}$. See Fig.~\ref{Fig_M}(a), $M_{\rm FEO}$ stays to be zero even for any $J_{xx}$ and $p_{z}$ and see Fig.~\ref{Fig_M}(b), $M_{\rm WFO}=0$ for $0\leq J_{xx}<1$ and the clear jump appears at $J_{xx}=1$, $M_{\rm WFO}=1$ for $1\leq J_{xx}$. Moreover, $p_z$-dependence of $M_{\rm SFO}$ for various $J_{xx}$'s is shown in Fig.~\ref{Fig_M}(c). We found that the strong symmetry fractionalization for the symmetry $W$ (which occurs in the pure SPT) gradually vanishes where the bit flip operator is no longer generated by applying the strong symmetry $W^*_{u}\otimes I_{\ell}$ to the state. This tendency is valid as long as $J_{xx}$-induced SSB does not occur.

The combination of the behaviors of $M_{\rm WFO}$ and $M_{\rm WFO}$ indicates that the state $|\rho^{D}\rangle\rangle$ behaves our expectation as shown in Fig.~\ref{Fig_PD}(b). Indeed, in $0\leq J_{xx}<1$ and $p_{z}>0$, the weak symmetry fractionalization as in Eq.~(\ref{WSF_eq}) occurs, but the strong one for the symmetry $W$ vanishes. This is the demonstration of the symmetry fractionalization of the cluster ASPT. In addition, we give the support results in Appendix F, where we numerically observe the spatial distribution of the edge bit and demonstrate the bit-flip through the symmetry fractionalization numerically.

{\it Operator space mutual information for doubled system.---}
Finally, we observe a mutual information between edges through 
the operator space entanglement entropy (OSEE) \cite{Prosen2007,Pizorn2009,Znidaric2008,Nieuwenburg2014,Nieuwenburg2018} for the decohered density matrix. 
This can be efficiently calculated in the Choi map, corresponding to the calculation of the subsystem entanglement entropy for the Choi state $|\rho^{D}_{\alpha\beta}\rangle\rangle$ on the doubled system (ladder system).
The subsystem entanglement entropy on the doubled system (ladder geometry) corresponds to the OSEE of the same subsystem, defined as
$
S^{\rm OS}(X)=-{\rm Tr}[{\tilde \rho}_X \log {{\tilde \rho}_X}]$
,
where ${\tilde \rho}_X$ is a reduced density matrix of the subsystem $X$ obtained from the density matrix of the renormalized Choi state ${\tilde \rho}\equiv |{\tilde \rho}^D\rangle\rangle\langle\langle {\tilde \rho}^D|$ with $|{\tilde \rho}^D\rangle\rangle=|\rho^D\rangle\rangle/\sqrt{\langle\langle \rho^D|\rho^D\rangle\rangle}$. 
Then, we focus on the operator space mutual information (OSMI) given by 
$
I^{\rm OS}(A,B)=S^{\rm OS}(A)+S^{\rm OS}(B)-S^{\rm OS}(AB)$, 
where the subsystem $A(B)$ is the left(right) edge-sites on the bra and ket Hilbert space, $(0,u)$, $(0,\ell) \in A$ and $(L-1,u)$ and $(L-1,\ell) \in B$. The detailed explanation of the OSEE and OSMI is shown in Appendix C. Note that it is believed that the OSEE and OSMI are the quantities characterizing the correlations between $A$ and $B$ subsystems \cite{Prosen2007,Nieuwenburg2018}.

We can suddenly observe $I^{\rm OS}(A,B)$ for two typical limiting cases as follows. 
As an initial state 
we prepare the Bell-pair edge states of the pure cluster SPT, $|\psi^B_0\rangle=[|0_A0_B\rangle_{LR}+|1_A1_B\rangle_{LR}]/\sqrt{2}$ where the bulk part is omitted and the state is approximately an edge state satisfying $(Z_0)[(Z_{L-1})]|\alpha_A\beta_B\rangle_{LR}=(-1)^{\alpha}[(-1)^{\beta}]|\alpha_A\beta_B\rangle_{LR}$ (exact for $J_{xx}=0$). The Choi map of the density matrix of the SPT with edge states can be regarded as two Bell-pair edge states on each upper and lower chain in the doubled system. 
Then, for $J_{xx}=0$, the OSMI picks up $I^{\rm OS}(A,B)=2\ln 2$, corresponding to the quantum correlation from two Bell-pairs between the edges (See Appendix D). 
We directly consider another limit $p_z=1/2$ from the $p_{z}=0$ case (See Appendix E). The $p_z=0$ state reduces a classical mixture given by $|\rho^D_{p_z=1/2}\rangle\rangle=[|\rho^c_{00}\rangle\rangle+|\rho^c_{11}\rangle\rangle]/2$ ($\rho^D_{p_z=1/2}=[\rho^c_{00}+\rho^c_{11}]/2$). 
The application of the decoherence $\hat{\mathcal{E}}^{Z}_{e}$ leads to the breakdown of one Bell pair between the edge from the initial state and the bulk ASPT. 
However, even for the classical mixture state, although one initial Bell pair breaks, the correlation between the left-right edges survives as estimated by the OSMI, taking $I^{\rm OS}(A,B)=\ln 2$. 
Its simple explanations for $I^{\rm OS}(A,B)=\ln 2$ based on the Choi mapping is shown in Appendix D.
\begin{figure}[t]
\begin{center} 
\includegraphics[width=7cm]{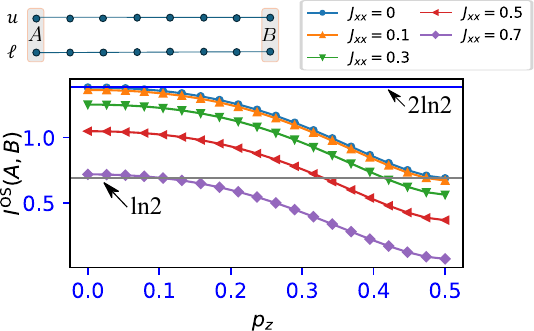}  
\end{center} 
\caption{$p_z$-dependence of the OSMI. 
In the upper panel, the partition of the entanglement entropy for the doubled system (ladder geometry). The subsystem $A$ and $B$ includes the edge-sites of the upper and lower Hilbert space. 
We set $L=23$ ($23\times 2$-sites ladder). The system size dependence for $J_{xx}=0$ is shown in Appendix F. 
}
\label{Fig_MI}
\end{figure}

Based on the above, we numerically verify the properties of the correlation between edges discussed above and observe its interpolated behavior and then how robust the correlation picture is for a finite $J_{xx}$-term. Technically, the initial SPT with Bell-pair edges can be created by adding infinitesimal effective edge interactions $-\epsilon (Z_0Z_{L}+X_0Z_1Z_{L-2}X_{L-1})$ to the Hamiltonian $H_c$. 
The behavior of the OSMI is shown in Fig.~\ref{Fig_MI}. It is confirmed that for $J_{xx}=0$ case, the initial quantum correlation ($2\ln 2$) between the edges decays as increasing $p_{z}=1/2$. The initial two Bell pairs decays and but the partial correlation between edge bits remains ($\ln 2$) even for $p_z\to 1/2$. This surely shows the presence of the edge correlation even for this ASPT.
For a finite $J_{xx}$, since the edge bits are deformed and the initial two Bell pairs of edges are not perfectly produced, the value of the OSMI entirely decreases but as increasing $p_z$, the OSMI decreases. This indicates that even under finite perturbations of $J_{xx}$-terms, initially the Bell-pair edge is produced and these are broken down but some correlation between edges remains even for maximal decoherence as far as the SSB does not occur.

{\it Conclusion.---} 
This work elucidated the symmetry fractionalization of the ASPT in the cluster model under decoherence and the behavior of the edge bits. 
We numerically demonstrated the behavior of the edge bits based on the Choi mapping.
By varying the strength of decoherence and adding a perturbation, how the symmetry fractionalization and edge-bits behave has been observed beyond the abstract level \cite{Ma_PRXQuantum6}.   
We also verified that the edge bit of the ASPT possesses an edge correlation originating from the quantum correlation between the edges in the pure SPT and clarified the flow of the correlations between edge bits as forming the ASPT by decoherences. This work represents a first step toward concretely establishing the bulk–edge correspondence in ASPTs and exploring its possible varieties. 


\section*{Acknowledgements}
This work is supported by JSPS KAKENHI: JP23K13026(Y.K.). 

\section*{Data availability}
The data that support the findings of this study are available from the authors upon reasonable request.
\bibliography{ref}

\clearpage
\renewcommand{\thesection}{A\arabic{section}} 
\renewcommand{\theequation}{A\arabic{equation}}
\renewcommand{\thefigure}{A\arabic{figure}}
\setcounter{figure}{0}
\widetext
\appendix
\section*{Appendix}
\section{A. Strong and weak symmetries}
Two types of notions of symmetries can be introduced in mixed states, namely, strong and weak symmetries \cite{Buca2012,groot2022}.
This work picks up the local unitary $Z_2$ symmetry version. 
The group is constituted by the generator set $\{\hat{1},U_{Z_2}\}$ where $U^2_{Z_2}=\hat{1}$. 

The strong symmetry of a density matrix is defined as 
$$
U_{Z_2}\rho=e^{i\theta}\rho,\;\;\; \rho U^{\dagger}_{Z_2}=e^{-i\theta}\rho,
$$ 
where $\rho$ is a mixed state and $\theta$ is a global phase factor. 

We further introduce the weak-symmetry condition defined as  
$$
U_{Z_2}\rho U^\dagger_{Z_2} = \rho.
$$ 
The symmetry is satisfied on the average level \cite{ma2024}, in the sense that the symmetry is satisfied after taking the ensemble average in the density matrix.

The notion of the strong and weak symmetry conditions are further defined on quantum channels. 
Genenrally, the quantum channel including decoherences can be described by the Kraus operator form \cite{Nielsen2011},
$\mathcal{E}(\rho)=\sum^{N-1}_{\ell=0}K_{\ell} \rho K^\dagger_{\ell}
$, where $\{K_{\ell}\}$ are a set of Kraus operators satisfying $\sum^{N-1}_{\ell=0} K^\dagger_{\ell} K_{\ell}=\hat{I}$ with $\hat{I}$ being the identity operation. 
The quantum channel $\mathcal{E}$ induces changes in mixed states. 
Here, the strong $Z_2$-symmetry condition on the channel is given as 
$K_{\ell}U_{Z_2}=e^{i\theta} U_{Z_2} K_{\ell}$ for any $\ell$. 
On the other hand, the weak symmetry condition on the channel is expressed as 
$
U_{Z_2}\biggl[\sum_{\ell}K_{\ell} \rho K^\dagger_{\ell}\biggr]U^\dagger_{Z_2}=\mathcal{E}(\rho)$.
This condition does not require that each Kraus operator commutes with non-trivial generator $U_{Z_2}$. 

\subsection*{B. Choi representation for decoherence channels}
The Choi representation \cite{Choi1975} makes a density matrix the supervector, namely the Choi map mentioned in the main text. 
In this formalism, quantum decoherences expressed in terms of the Kraus operator can be represented as a non-unitary operator to the Choi map. The decoherence can be regarded as an operator application to the Choi map. 

If for a density matrix, the decoherence is written as ~\cite{Nielsen2011,lidar2020}$$\mathcal{E}[\rho]=\sum^{M-1}_{\alpha=0}K_{\alpha}\rho K^\dagger_{\alpha},
$$ 
where $K_\alpha$'s are Kraus operators satisfying $\sum^{M-1}_{\alpha=0}K_{\alpha}K^{\dagger}_{\alpha}=I$, then
The decoherence is also transformed as the following operator ~\cite{Lee2025} 
$
\mathcal{E}\longrightarrow \hat{\mathcal{E}}=\sum^{M-1}_{\alpha=0}K^*_{\alpha,u}\otimes K_{\alpha,\ell}$. 
The original Kraus operators become the super operator acting on the upper and lower Hilbert spaces, respectively.  
For the Choi map $|\rho\rangle\rangle$, the operator $\hat{\mathcal{E}}$ acts as 
$
\hat{\mathcal{E}}|\rho\rangle\rangle$. 
This application changes the mixed state into a different mixed state.
Although the decoherence operator is not a unitary map in general cases, the operation is a completely positive trace-preserving map.  
The application of the decoherence operator generally changes the norm of the Choi map, which is nothing but the purity of the original mixed state.

\subsection*{C. Operator space entanglement entropy}
In this study, we introduce the OSEE employed in this work. The Choi map of the renormalized decohered density matrix is generally written as $|\tilde{\rho}^D\rangle\rangle=\frac{1}{\sqrt{\Tr[(\rho^D)^2]}}\sum_{i,j}|i\rangle_u\otimes \rho^D|j\rangle_{\ell}$, where $|i\rangle_{u}$ and $|j\rangle_{\ell}$ are a basis of the upper and lower Hilbert spaces. We here perform the renormalization because it provides a clear interpretation of the subsequent observable as a well-defined correlation measure. We consider the density matrix for the state $|\tilde{\rho}^D\rangle\rangle$ as 
$
\tilde{\rho}^D\equiv|\tilde{\rho}^D\rangle\rangle \langle\langle\tilde{\rho}^D|
$.
We partition the total system into two subsystems, $A$ and $B$, and introduce the same spatial-partition for each upper and lower basis as $|i\rangle_{u}\to |(i_A,i_B)\rangle_u$ and $|j\rangle_{\ell}\to |(j_A,j_B)\rangle_{\ell}$. 
For this partition, we consider the entanglement entropy for the subsystem $A$. Here, we consider the reduced density matrix for the subsystem $A$, which is given as
\begin{eqnarray}
&&\tilde{\rho}_A=\Tr[\tilde{\rho}^D]=\sum_{(i_A,i'_A),(j_A,j'_A)}\tilde{\rho}'_{A,(i_A,j_A),(i'_A,j'_A)}[|i_A\rangle_u|j_A\rangle_\ell {}_{u}\langle i'_A|{}_{\ell}\langle j'_A|],\\
&&\tilde{\rho}'_{A,(i_A,j_A),(i'_A,j'_A)}\equiv\sum_{i_B,j_B}\frac{1}{\Tr[(\rho^{D})^2]}\rho^D_{(i_A,i_B),(j_A,j_B)}\rho^{D*}_{(i'_A,i_B),(j'_A,j_B)}.
\end{eqnarray}
The formulation of this reduced density matrix $\tilde{\rho}'_{A,(i_A,j_A),(i'_A,j'_A)}$ just corresponds to the formulation of the reduced density matrix in the operator-space entanglement entropy (OSEE)\cite{Prosen2007}. That is, in the formulation of the OSEE, we just pick up the density matrix $\rho^D$ as the target operator. A similar pickup for the choice of operator has already been tested in \cite{Znidaric2008,Nieuwenburg2014} . 
Based on this correspondence, the entanglement entropy for the subsystem $A$ of the state $|\tilde{\rho}^D\rangle\rangle$ is given by
\begin{eqnarray}
S^{\rm OS}=-\Tr[\tilde{\rho}'_{A}\ln \tilde{\rho}'_{A}]
\end{eqnarray}
corresponds to the OSEE for the density matrix $\tilde{\rho}^D$. It is used as a correlation measure between $A$ and $B$ subsystems. Many previous studies have succeeded in characterizing correlations between subsystems by using the OSEE \cite{Prosen2007,Znidaric2008,Pizorn2009,Nieuwenburg2014,Nieuwenburg2018}.

\subsection*{D. Correlations between edges for representative limits}
We analytically observe the OSEE for two representative state. 
In what follows, the subsystem $A(B)$ represents the left(right) edge-sites on the bra and ket Hilbert space, $(0,u)$, $(0,\ell) \in A$ and $(L-1,u)$ and $(L-1,\ell) \in B$.\\

\underline{(I) The state double cluster SPT with Bell-pair edges:} 
For $p_z=0$ and $J_{xx}=0$, the Choi map of the density matrix for the pure cluster SPT effectively is given by 
\begin{eqnarray}
|\Psi_{p_z=0}\rangle\rangle=\frac{1}{2}\biggr[|(0_A0_B)\rangle_{u}|\psi^b_{00}\rangle_{u}+|1_A1_B\rangle_{u}|\psi^b_{11}\rangle_{u}\biggl] \otimes \biggr[|0_A0_B\rangle_{\ell}|\psi^b_{00}\rangle_{\ell}+|1_A1_B\rangle_{\ell}|\psi^b_{11}\rangle_{\ell}\biggl], 
\label{p=0_SD}
\end{eqnarray}
where $|(\alpha_A\beta_B)\rangle_{u(\ell)}$ represents the left and right ledge states for the upper(lower) space of the $A$ and $B$ subsystems and $|\psi^b_{\alpha\beta}\rangle_{u(\ell)}$ is the bulk part of the state depending on the left and right edge states labeled by $\alpha$ and $\beta$.
This description can correctly capture correlations between the edges in the system. 
For this state $|\Psi_{p_z=0}\rangle\rangle$, we can directly calculate the OSEE $S^{\rm OS}(A)$. The direct calculation is $S^{\rm OS}(A)=2\ln 2$. Similarly, $S^{\rm OS}(B)=2\ln 2$ and $S^{\rm OS}(AB)=2\ln 2$. Thus, $I^{\rm OS}(A,B)=2\ln 2$.  
As an additional test, we observe negativity defined as \cite{peres1996,Horodecki_family}
\begin{eqnarray}
\mathcal{N}_X\equiv \ln|\rho^{\Gamma_X}|_1,
\end{eqnarray}
where $X$ is a subsystem, $|\cdot|_1$ represents the trace norm, and $\Gamma_X$ is a partial transpose operation for $X$-subsystem. It is believed that the negativity can extract only quantum correlation \cite{Werner1989,peres1996,Horodecki_family}. On the same partition, we calculate the mutual negativity defined as $I^N(A,B)\equiv \mathcal{N}_{A}+\mathcal{N}_B-\mathcal{N}_{AB}$. We obtain $I^N(A,B)=2\ln 2$. From this result, we expect that the result, $I^{\rm OS}(A,B)=2\ln 2$ indicates that the quantum correlation is $2\ln 2$. 
This correspondence between the mutual information and negativity is natural because we focus on the entanglement entropy for the purified Choi map. 
Also, if the form of Eq.~(\ref{p=0_SD}) just corresponds to the presence of two Bell-pair edges on both upper and lower Hilbert spaces. Thus, the type of the correlation of $I^{\rm OS}(A,B)=2\ln 2$ is quantum.
\\

\underline{(II) $p_z=1/2$ fixed cluster ASPT:}
We next observe $p_z=1/2$ fixed cluster ASPT. 
The form of the state can be deduced from the state $|\Psi_{p_z=0}\rangle\rangle$. The mechanism is explained in Appendix E. The Choi map of the $p_z=1/2$ fixed cluster ASPT is
\begin{eqnarray}
|\Psi_{p_z=1/2}\rangle\rangle=\frac{1}{\sqrt{2}}\biggr[|0_A0_B\rangle_{u}|0_A0_B\rangle_{\ell}|\psi^b_{00}\rangle_{u\ell} +|1_A1_B\rangle_{u}|1_A1_B\rangle_{\ell}|\psi^b_{11}\rangle_{u\ell}\biggl].
\end{eqnarray}
Here, $|\psi^b_{\alpha\beta}\rangle_{u(\ell)}$ is the upper-lower connected bulk part of the state depending on the left and right edge states labeled by $\alpha$ and $\beta$ the reason of which is that the bulk is entangled between the upper and lower part due to the decoherence.
The above Choi map is already renormalized.

Due to the decoherence there is no off-diagonal interference term such as $|0_L0_R\rangle_{u}|1_L1_R\rangle_{\ell}$ on the local left-right edge basis. 
This suppresses the correlation between edges. 
From the above Choi state $|\Psi_{p_z=1/2}\rangle\rangle$, the OSEE is analytically calculated. The result is $S^{\rm OS}(A)=\ln 2$. Similarly, $S^{\rm OS}(B)= S^{\rm OS}(AB)=\ln 2$. Thus, $I^{\rm OS}(A,B)=\ln 2$. Moreover, as an additional test, we calculate the mutual negativity $I^N(A,B)\equiv \mathcal{N}_{A}+\mathcal{N}_B-\mathcal{N}_{AB}$. We obtain $I^N(A,B)=\ln 2$ corresponding to $I^{\rm OS}(A,B)=\ln 2$. 
Based on the Choi map, even for the maximally decohered ASPT, the correlation between edges stays to be a finite value, that is, the initial Bell-pair correlation partially remains.

\subsection*{E. Suppression of off-diagonal edge state by decoherences and appearance of ASPT}
We here discuss the suppression of the off-diagonal term, such as $|0_A0_B\rangle_{u}|1_A1_B\rangle_{\ell}$, that appeared in the previous appendix.
This suppression is induced by the local odd-site decoherence of $\mathcal{E}^{Z}_{e}$. 
In the Choi formalism, this decoherence is regarded as an interaction between the upper site and lower site in the doubled system \cite{Haegeman2015,Orito2025}. That is, concretely
\begin{eqnarray}
\hat{\mathcal{E}}^{Z}_{e}&=&\prod_{j\in even}\biggr[(1-p_{z})\hat{I}_{j,u}^* \otimes \hat{I}_{j,\ell} +p_{z}Z_{j,u}^*\otimes Z_{j,\ell}\biggl]\propto e^{-\tau(p_z) \hat{h}_{zz}},\\
\hat{h}_{zz}&\equiv& -\sum_{j\in odd}Z_{j,u}Z_{j,\ell}
\end{eqnarray}
where $\tau(p_z)=\tanh^{-1}[{p_{z}/(1-p_{z})}]$ and $\hat{h}_{zz}$ is an effective interaction between the upper and lower chain. This decoherence operation acts like the imaginary time evolution operator to the initial state $|\Psi_{p_z=0}\rangle\rangle$. Then, since the sign of the interaction $\hat{h}_{zz}$ is negative, the off-diagonal terms in the state $|\Psi_{p_z=0}\rangle\rangle$ are energetically suppressed as increasing $p_z$. The upper and lower edge states tend to align. This is a suppression mechanism of the quantum correlation between edges on the Choi doubled system.
In principle, once the decoherence $\hat{\mathcal{E}}^{Z}_{e}$ is switched on, the mechanism is effective even for any finite $p_z$. Thus, the quantum correlation between edges in $|\Psi_{p_z=0}\rangle\rangle$ is gradually swept out since the interaction disfavors the off-diagonal components of the initial state.
As these off-diagonal components diminish, the upper and lower edges effectively align, placing the system in the ASPT fixed-point sector in Fig.~\ref{Fig_PD}. 

\subsection*{F. System size dependence}
\begin{figure}[h]
\begin{center} 
\includegraphics[width=13.5cm]{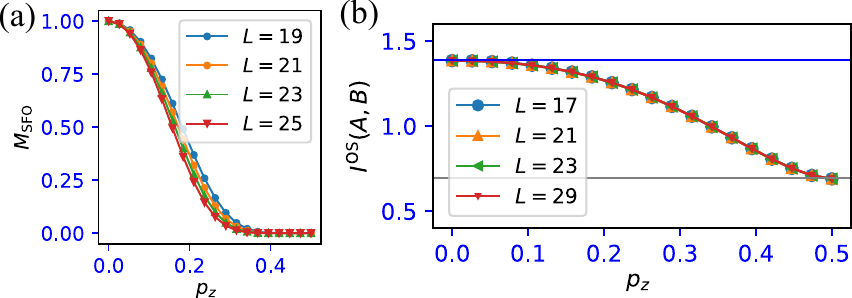}  
\end{center} 
\caption{(a) System-size dependence of the symmetry fractionalization order parameter $M_{\rm SFO}$ with $J_{xx}=0$. (b) System-size dependence of the OSMI $I^{\rm OS}(A,B)$ with $J_{xx}=0$. The blue and gley lines represent $I^{\rm OS}(A,B)=2\ln 2$ and $\ln 2$, respectively. The simulation is carried out on the ladder geometry where the total number of sites is $2L$.
}
\label{Fig_SSD}
\end{figure}

We show the typical system-size dependence of the symmetry fractionalization order parameter $M_{\rm SFO}$ and the OSMI. The data is shown in Fig.~\ref{Fig_SSD}. The system-size dependence of $M_{\rm SFO}$ at $J_{xx}=0$ appears only slightly in the middle of $p_{z}$ as shown in Fig.~\ref{Fig_SSD}(a). On the other hand, the OSMI at $J_{xx}=0$ does not exhibit the dependence at all as shown in Fig.~\ref{Fig_SSD}(b). From these numerical results, the system size we took is sufficiently large to suppress finite-size effects. For other parameter sets and the physical quantities $M_{\rm FEO}$ and $M_{\rm WFO}$, our simulation reaches sufficient system size to exhibit almost no system size dependence.

\subsection*{G. Spatial distribution of edge bit and observing symmetry fractionalization}
To support the discussion about the presence of the edge bits in the target system and the presence of the symmetry fractionalization for the weak symmetry operation by the generator $W$. 
Here, we numerically show the presence of edge bit by plotting the spatial distribution of the spin $Z_j$ expectation value, and also observe the symmetry fractionalization for the weak symmetry.

In the doubled Hilbert space formalism, from the Choi map of the state, the expectation value of $Z_j$ can be calculated as \cite{Orito2025}
$$
\langle Z_j\rangle=\frac{\langle\langle {\bf 1}|Z_{i,u}|\rho_D\rangle\rangle}{\langle\langle {\bf 1}|\rho_D\rangle\rangle},
$$
where $|{\bf 1}\rangle\rangle\equiv \displaystyle{\frac{1}{2^{3L/2}}\prod^{L-1}_{j=0}|t\rangle_j}$ with $|t\rangle_j=|\uparrow_u\uparrow_{\ell}\rangle_j+|\downarrow_u\downarrow_{\ell}\rangle_j$ and the corresponding quantity in the original physical Hilbert space is ${\rm Tr}[\rho_D Z_{j}]$.  

We also observe how the weak symmetry operator acts on a state with a classical bit fixed on $Z_0=\alpha$ and $Z_{L}=\beta$
and if the symmetry fractionalization occurs, 
the symmetry operator changes the boundary edge flip operators acting around edges, and further flips the left and right edge bits as  
\begin{eqnarray}
&&(W^*_u\otimes W_{\ell})|\rho^D_{\alpha\beta}\rangle\rangle
=(Q^{(p)}_{L,u}Q^{(p)}_{R,u})(Q^{(p)}_{L,\ell}Q^{(p)}_{R,\ell})|\rho^D_{\alpha\beta}\rangle\rangle=|\rho^D_{\bar{\alpha}\bar{\beta}}\rangle\rangle.
\end{eqnarray}
We observe whether or not such a bit flip action for pure and mixed states occurs through the symmetry fractionalization. 

Let us show the numerics. Here, as an initial state, we create the state by adding the edge magnetic field $-\epsilon (Z_0+Z_{L})$ to the cluster Hamiltonian $H_c$ by the DMRG. By its procedure, an initial $\alpha=\beta=1$ edge state is prepared. Then, we apply the decoherence $\mathcal{E}^Z_e$ with $p_z=1/2$ to the initial state and observe $\langle Z_j\rangle$. 

\begin{figure}[t]
\begin{center} 
\includegraphics[width=18cm]{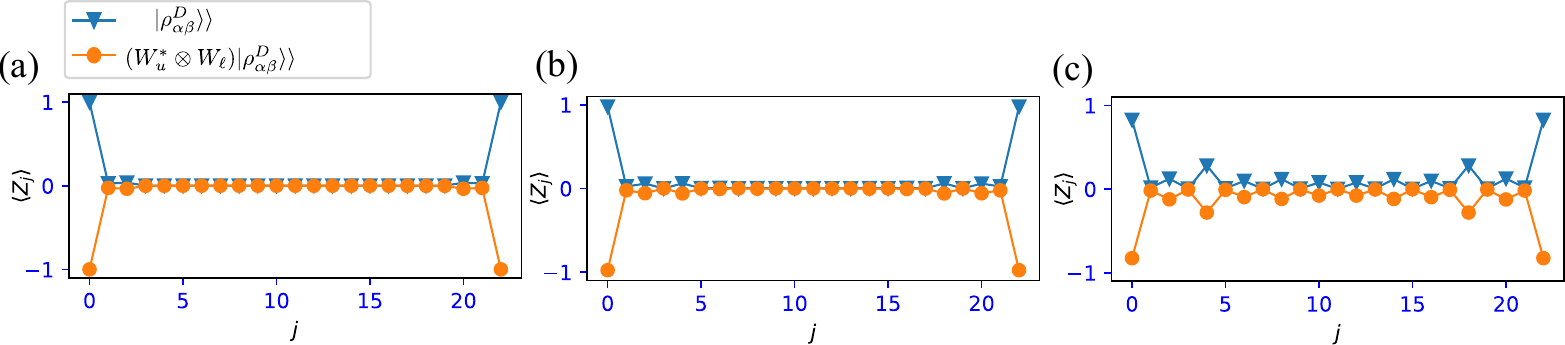}  
\end{center} 
\caption{Spacial distribution of the expectation value of $Z_j$. (a) $p_z=1/2$ and $J_{xx}=0$. (b) $p_z=1/2$ and $J_{xx}=0.4$. (c) $p_z=1/2$ and $J_{xx}=0.8$.
For all case, we set $L=23$. 
}
\label{Fig_Z_dis}
\end{figure}

For typical $J_{xx}$'s, the results of $\langle Z_j\rangle$ is shown in Fig.~\ref{Fig_Z_dis}. The blue point data represents $\langle Z_j\rangle$ without applying $(W_u\otimes W_{\ell})$ to the decohered state. We surely confirm the presence of the local bit around both edges. 
In particular, for $J_{xx}=0$ and $0.4$ data [Fig.~\ref{Fig_Z_dis} (a) and Fig.~\ref{Fig_Z_dis} (b)], exact localized bit appears where $\langle Z_{0(L)}\rangle=+1$. For the strong perturbation case $J_{xx}=0.8$ [Fig.~\ref{Fig_Z_dis}(c)], we observe deformed edge bits where the bits, of course, still localize around edges [the large values of $\langle Z_j\rangle$ stay on the edges], but there is a leak into the bulk. Next, we see the case of the after applying the weak symmetry operator $(W^*_u\otimes W_{\ell})$ to the decohered state. The cases of $\langle Z_j\rangle$ are presented as orange dot plot data in Fig.~\ref{Fig_Z_dis}. We verify that the pattern of $\langle Z_j\rangle$ is completely flipped, that is, the symmetry operation is surely fractionalized, and the emergence edge operators flip the edge bits. Especially, note that for the strong perturbation case $J_{xx}=0.8$ [Fig.~\ref{Fig_Z_dis}(c)], the deformed edge distribution indicates that the edge operator is no longer exact $Z_0$ and $Z_{L}$, however, accompanying this deformation, one can also observe that the edge-flip operator arising from symmetry fractionalization is itself deformed, and that it flips the corresponding deformed edge bit.

\end{document}